\documentstyle[prd,aps,preprint,psfig]{revtex}
\begin{document}
\newcommand{\gsim}{ \mathop{}_{\textstyle \sim}^{\textstyle >} }
\newcommand{\lsim}{ \mathop{}_{\textstyle \sim}^{\textstyle <} }
\newcommand{\vev}[1]{ \left\langle {#1} \right\rangle }
\newcommand{\bra}[1]{ \langle {#1} | }
\newcommand{\ket}[1]{ | {#1} \rangle }
\newcommand{\EV}{ {\rm eV} }
\newcommand{\KEV}{ {\rm keV} }
\newcommand{\MEV}{ {\rm MeV} }
\newcommand{\GEV}{ {\rm GeV} }
\newcommand{\TEV}{ {\rm TeV} }

\tighten
\draft
\title{${\cal N}=2$ Supersymmetry in a Hybrid Inflation Model}
\author{T. Watari$^a$ and T. Yanagida$^{a,b}$}
\address{$^a$ Department of Physics, University of Tokyo, Tokyo 113-0033,
Japan \\ and \\
$^b$ Research Center for the Early Universe, University of
Tokyo, Tokyo, 113-0033, Japan}

\date{\today}

\maketitle

\begin{abstract}
The slow roll inflation requires an extremely flat inflaton potential.
The supersymmetry (SUSY) is not only motivated from the gauge hierarchy
 problem, but also from stabilizing that flatness of the
 inflaton potential against radiative corrections. However,  it has 
 been known that the Planck suppressed higher order terms in the K\"ahler potential 
 receive large radiative corrections loosing the required flatness in the ${\cal N}=1$ supergravity.  
We propose to impose a global ${\cal N}=2$ SUSY on the inflaton
 sector. What we find is that the ${\cal N}=2$ SUSY Abelian gauge 
theory is exactly the same as the desired hybrid inflation
 model. The flat potential at the tree level is not our choice of
 parameters but a result of the symmetry.
We further introduce a cut-off scale of the theory which is lower than the Planck
 scale. This lower cut-off scale suppresses the
 supergravity loop corrections to the flat inflaton potential.
\end{abstract}

\clearpage


Inflationary universe is widely believed as the standard theory of 
modern cosmology \cite{Linde}, since there has been found, so far, 
no alternative to solve the fundamental problems in cosmology, i.e. 
the horizon and the flatness problems. The inflationary cosmology leads
to a very flat universe ($\Omega \simeq 1$) at the present time, and this
remarkable prediction has been strongly 
supported by recent observations on cosmic microwave background 
radiations (CMBR) by Boomerang \cite{boomerang} and Maxima
\cite{maxima}. The inflationary 
universe requires a scalar field called as inflaton $\varphi$ which
should have necessarily 
a very flat potential to generate the inflation in the early
universe. However, this flat potential is subject to radiative
corrections and it receives easily a large deformation loosing the
required flatness. Supersymmetry (SUSY) is an interesting theory which may 
protect the flat potential for the inflaton $\varphi$ from 
having large radiative corrections. In fact, the superpotential is 
completely stable against the radiative corrections due to the 
nonrenormalization theorem \cite{nonrenormal}. However, the K\"{a}hler 
potential receives large radiative corrections even in the SUSY theories 
\cite{Gaillard} and we need a fine-tuning of various
parameters in supergravity to maintain the flatness of inflaton    
potential at the quantum level \cite{Stewart}.

A possible solution to the above problem is to introduce a cut-off scale 
$M_*$ much below the Planck scale $M_{\rm Planck}\simeq 2\times 10^{18}$ GeV
in order to suppress the unwanted radiative corrections in the K\"{a}hler
potential for inflaton $\varphi$. We adopt the cut-off scale $M_* \ll
M_{\rm Planck}$ throughout this paper. However, the introduction of the 
cut-off raises a new serious problem; nonrenormalizable operators of 
the inflaton field $\varphi$ are suppressed by inverse powers of the
cut-off scale $M_{*}$ which easily destroys the
flatness of the scalar potential $V$ because of $M_{*} \ll M_{\rm Planck}$.
The situation becomes even worse.

In this paper, we propose to use a ${\cal N}=2$ SUSY to control the 
tree-level K\"{a}hler and superpotentials for the inflaton and construct an explicit model 
(the hybrid inflation model) which may avoid the above mentioned fine-tuning
problem. In this model the form of K\"{a}hler and superpotentials are 
fixed by the global ${\cal N}=2$ SUSY and the additional R-symmetry. In
particular, we show that the tree-level K\"{a}hler potential has necessarily the
minimal form and hence it is independent of the
cut-off scale $M_*$. Thus, the flat potential is guaranteed by the
symmetries. The gravity interactions, however, break explicitly the 
global ${\cal N}=2$ SUSY, since we impose only the ${\cal N}=1$ 
supergravity in the full theory.  We calculate radiative corrections 
from the supergravity sector and find that they are sufficiently
suppressed owing
to the small cut-off scale $M_*$. Thus, we consider that our hybrid 
inflation model based on the global  ${\cal N}=2$ SUSY is perfectly natural.

Before describing our explicit model we first discuss our basic
assumptions. First of all, we adopt the framework of ${\cal N}=1$
supergravity for the full theory whose scalar potential is determined by 
two fundamental potentials, i.e. K\"{a}hler potential 
$K(\Phi ^{\dagger \bar{i}}, \Phi ^i)$ and superpotential $W(\Phi ^i)$, 
where $\Phi ^i$ are chiral superfields. 
The scalar potential is given by
\begin{eqnarray}
V={\rm exp}\left( \frac{K(\phi ^{\dagger}, \phi )}{M^2_{\rm Planck}}\right)
    \left[ \left(W_i + \frac{K_i}{M^2_{\rm Planck}} W \right)
           \left(W^\dagger_{\bar{j}} + \frac{K_{\bar{j}}}{M^2_{\rm Planck}}W^\dagger \right) K^{\bar{j}i} 
        - 3 \left|\frac{W(\phi )}{M_{\rm Planck}}\right|^2 \right],
\label{eq:tree-sugra}
\end{eqnarray}
%
where 
\begin{eqnarray}
  W_i = \frac{\partial W(\phi)}{\partial \phi^i}, & \qquad 
& W_{\bar{j}}^\dagger = \frac{\partial W^\dagger(\phi^\dagger)}
                             {\partial \phi^{\dagger \bar{j}}}, \\
  K_i = \frac{\partial K(\phi,\phi^\dagger)}{\partial \phi^i}, & \qquad
& K_{\bar{j}} = \frac{\partial K(\phi,\phi^\dagger)}
                     {\partial \phi^{\dagger \bar{j}}},   \\
  K_{i \bar{j}} = \frac{\partial^2 K(\phi,\phi^\dagger)}
                       {\partial \phi^i \partial \phi^{\dagger \bar{j}}}, 
& \qquad 
& K_{i \bar{j}}(\phi,\phi^\dagger) K^{\bar{j} k}(\phi,\phi^\dagger) 
   = \delta_i^{\;\; k}.
\end{eqnarray}
Here, $\phi^i, \phi^{\dagger \bar{j}}$ are scalar components of the
(anti-) chiral superfields $\Phi^i, \Phi^{\dagger \bar{j}}$. 
The first nontrivial assumption is that the K\"{a}hler potential $K$ and
superpotential $W$ do not depend on the Planck scale $M_{\rm Planck}$ 
but depend on the cut-off scale $M_*$.\footnote{
Here, we suppose that the K\"{a}hler potential $K$ has a fundamental meaning
rather than the $ -3 M^2_{\rm Planck} \exp( - K/(3M^2_{\rm Planck}))$. 
The former describes the metric of manifold of matter
superfields. Its geometrical meaning is clear. The
latter is a natural  
expression in the superspace Lagrangian of the supergravity\cite{Wess-Bagger}. 
} 
This means that the basic interactions for matter superfields $\Phi ^i$
are controlled only by the 
cut-off scale $M_*$ instead of the Planck scale $M_{\rm Planck}$. In 
particular, the metric of the scalar components $\phi ^i$ is given by 
the K\"{a}hler potential as
\begin{eqnarray}
{\cal L} = 
   K_{i \bar{j}}(\phi,\phi^\dagger) \partial _\mu \phi ^{\dagger \bar{j}}\partial ^\mu \phi ^i,
\end{eqnarray}
and hence the metric of the scalar fields $\phi ^i$ is controlled by the 
cut-off scale $M_*$. We stress that this is a necessary assumption to 
obtain the desired flat potential as we will see below.
It is beyond the scope of this paper to discuss the
underlying physics that may justify this assumption.

The second assumption is rather simple, which is that the Lagrangian for 
the inflaton sector possesses a global ${\cal N}=2$ SUSY in the limit of 
$M_{\rm Planck}= \infty$. This assumption implies that when one switches 
off the ${\cal N}=1$ supergravity interactions, the inflaton sector is 
completely decoupled from other sectors including the standard-model fields.
This assumption together with the previous one means that the field
theory described by the K\"{a}hler and superpotentials, $K$ and $W$, 
for the inflaton sector alone is invariant under the global ${\cal N}=2$ 
SUSY, since they are independent of the Planck scale $M_{\rm Planck}$. 

We now discuss the inflaton sector which possesses the global ${\cal N}=2$
SUSY. We consider a $U(1)$ gauge theory with a hypermultiplet
($\Psi$ and ${\bar \Psi}$). $\Psi$ and ${\bar \Psi}$ are charged under the 
$U(1)$ gauge symmetry with the charge $+1$ and $-1$, respectively. 

The general superpotential of the ${\cal N}=2$ $U(1)$ gauge theory are
\cite{Fayet} 
\begin{equation}
 W = \sqrt{2} ( \bar{\Psi}\Phi \Psi - \mu^2 \Phi),
\label{eq:super-1}
\end{equation}  
where $\Phi$ is the ${\cal N}=1$ chiral superfield which is a partner of
the ${\cal N}=1$ U(1) gauge vector multiplet in the ${\cal N}=2$ SUSY theory.
The first term is one of the ``gauge interactions'', and forms a ${\cal N}=2$ 
theory along with the ordinary ${\cal N}=1$ gauge interaction in the K\"{a}hler
potential. The second term is one of the ``Fayet-Iliopoulos term''.
This term $-\sqrt{2}\mu^2 \Phi |_{\theta^2} $, along with the ordinary Fayet-Ilipoulos 
$D$-term ${\cal L}_{\rm FI-D} = -\xi^2 D$, forms a $SU(2)_R$ singlet
${\cal L}_{\rm FI}= -(2{\rm Im}\mu^2,2{\rm Re}\mu^2,\xi^2) \cdot
(-\sqrt{2}{\rm Im}F, \sqrt{2}{\rm Re}F,D)^{T}$\cite{Fayet}. 
The $SU(2)_R$ symmetry is explicitly broken by the $SU(2)_R$ triplet
vacuum-expectation value 
$(2{\rm Im}\mu^2,2{\rm Re}\mu^2,\xi^2)$. However, this superpotential
(\ref{eq:super-1}) is invariant under the $U(1)_R$ symmetry, where the
$R$-charges of $\Phi ,\Psi$ and ${\bar \Psi}$ are 2,0 and 0,
respectively. We impose the $U(1)_R$ symmetry throughout this paper.  

The K\"{a}hler potential for $\Phi$
is determined from the prepotential ${\cal F}(\Phi,M_*)$ as\cite{Gates}
\begin{eqnarray}
K(\Phi, \Phi^\dagger, M_*) = \frac{1}{4\pi}{\rm Im}\left(\Phi ^{\dagger} \frac {\partial}{\partial \Phi}{\cal F}(\Phi , M_*) \right).
\label{eq:special-geomtr}
\end{eqnarray}
Here, the prepotential ${\cal F}(\Phi,M_*)$ is a holomorphic function of
the superfield $\Phi$ carrying R-charge 4 and hence the R-invariance requires
\begin{eqnarray}
{\cal F}(\Phi , M_*) = \frac{1}{2} \Phi ^2
              \left(\frac{\vartheta}{2\pi} + i \frac{4\pi}{g^2}\right),
\end{eqnarray}
which leads to the minimal K\"{a}hler potential for $\Phi$,
\begin{eqnarray}
K = \frac{1}{g^2} \Phi ^{\dagger}\Phi.
\label{eq:minimal-1}
\end{eqnarray}
Here, $g$ is the $U(1)$ gauge coupling constant. We should stress 
that the minimal K\"{a}hler potential for $\Phi$ is not our 
choice by hand, but rather a result of the symmetries, i.e. the ${\cal
N}=2$ SUSY and the R-invariance. This is a big merit of the ${\cal N}=2$ SUSY.
Notice that the K\"{a}hler potential of $\Phi$ is
independent not only of $M_{\rm Planck}$ but also of the cut-off scale
$M_*$. 

The renormalization of the chiral superfield $\Phi$ is necessary to
obtain the canonically normalized kinetic term,
\begin{equation}
K = \Phi^\dagger \Phi.
\label{eq:minimal-2}
\end{equation} 
Then, the superpotential (\ref{eq:super-1}) becomes
\begin{equation}
 W = \sqrt{2}g \Phi (\Psi \bar{\Psi} - \mu^2).
\label{eq:super-2}
\end{equation}
It is very remarkable that the superpotential (\ref{eq:super-2}) along
with the minimal K\"{a}hler potential (\ref{eq:minimal-2}) of
$\Phi$ is exactly the same as that in the hybrid inflation model
\cite{Lyth-Riotto,Linde-Riotto}, where one of the scalar component $\phi$ of the $\Phi$
superfield plays a role of the inflaton $\varphi$. 

On the other hand, the K\"{a}hler potential for hypermultiplet ($\Psi$
and ${\bar \Psi}$) contains non-renormalizable terms. Indeed the K\"{a}hler
manifold of hypermultiplet is a hyperK\"{a}hler manifold, and it is not
determined by the special geometry as in eq.(\ref{eq:special-geomtr}). The
hyperK\"{a}hler condition allows
\begin{equation}
 K(\Psi,\bar{\Psi},M_*) = \Psi^\dagger \Psi + \bar{\Psi}^\dagger \bar{\Psi} 
  + \frac{k}{4M_*^2}\left( (\Psi^\dagger \Psi)^2 - 4 (\Psi^\dagger \Psi \bar{\Psi}^\dagger \bar{\Psi}) + (\bar{\Psi}^\dagger \bar{\Psi})^2 \right) + \cdots ,
\label{eq:hyper-kahler}
\end{equation} 
where $k$ is a real constant
, and
ellipses represent higher order terms. 
We see that $\psi = \bar{\psi} = 0$ is at least a local minimum of the
scalar potential for $\psi$ and $\bar{\psi}$ during the inflation. Here,
$\psi$ and $\bar{\psi}$ are scalar components of the hypermultiplet
($\Psi$ and $\bar{\Psi}$). The origin $\psi=\bar{\psi}=0$ may not always 
be the absolute minimum of the scalar potential for $\psi$ and
$\bar{\psi}$. 
However, in the following analysis, we assume that the 
$\psi = \bar{\psi} = 0 $ is the 
absolute minimum\footnote{
This is not a fine-tuning of parameters in the K\"ahler potential 
(\ref{eq:hyper-kahler})}, and $\psi$ and $\bar{\psi}$ stay at the origin
$\psi = \bar{\psi} =0 $ during the inflation.


The tree-level K\"{a}hler potential for the inflaton
sector may have a deformation due to the quantum effects. If it is
too large we loose the flat potential for the inflaton $\varphi$ and a
sufficiently long inflation is not expected. 

First, we discuss the radiative corrections through the ${\cal N}=2$
SUSY interactions. At quantum level, the $U(1)_R$ symmetry is
broken by $U(1)_R$-$(U(1)~{\rm gauge})^2$ anomaly. Under the $U(1)_R$
transformation $\Psi(\theta) \rightarrow \Psi(e^{i\alpha}\theta)$, 
$\bar{\Psi}(\theta) \rightarrow \bar{\Psi}(e^{i\alpha}\theta)$ and 
$\Phi(\theta) \rightarrow e^{-2i\alpha}\Phi(e^{i\alpha}\theta)$, the
Lagrangian is not invariant but 
\begin{equation}
 \delta{\cal L} = \frac{-4\alpha}{32\pi^2}F_{\mu\nu}\tilde{F}^{\mu\nu}.
\end{equation}
This variation of the Lagrangian, however, can be compensated by the
variation according to the transformation of the $\vartheta$ parameter
in the Lagrangian as
\begin{equation}
 \frac{\vartheta}{32\pi^2}F_{\mu\nu}\tilde{F}^{\mu\nu} \rightarrow
 \frac{\vartheta+4\alpha}{32\pi^2}F_{\mu\nu}\tilde{F}^{\mu\nu} .
\end{equation}
Then, we have a symmetry ``$U(1)_R$'' even at the quantum level
\cite{spurion-symetry}, under which the gauge coupling
spurion transforms as 
\begin{equation}
\frac{\vartheta}{2\pi} + i \frac{4\pi}{g^2} \rightarrow
\frac{\vartheta + 4\alpha}{2\pi} + i \frac{4\pi}{g^2},\label{eq:spurious-trans}
\end{equation}
along with the ordinary $U(1)_R$ transformation of the $\Psi,\bar{\Psi}$ 
and $\Phi$ with R-charge 0,0 and 2.

Whole radiative corrections to the K\"ahler 
potential of $\Phi$ are described by the deformation of the
prepotential. The deformation allowed by the ``$U(1)_R$'' symmetry is
\begin{equation}
 {\cal F}(\Phi,M_*) = \frac{1}{2} \Phi^2 \left[ \left(\frac{\vartheta}{2\pi}+i\frac{4\pi}{g^2(M_*)} - i \frac{2}{2\pi} \ln \left(\frac{\Phi}{M_*}\right)\right) + c_1 e^{\left(-\frac{8\pi^2}{g^2(M_*)} + i \vartheta\right)}\left(\frac{\Phi}{M_*}\right)^2 + \cdots \right].
\label{eq:prepote-deform}
\end{equation}
Note that the $M_*^{-2} \exp \left(-\frac{8\pi^2}{g^2(M_*)} + i
\vartheta\right) $ is charged under the ``$U(1)_R $'' with the charge
$-4$ (see eq.(\ref{eq:spurious-trans})).
The first term in the square bracket $\left[ \right.$  $ \left. \right]$ 
corresponds to the 1-loop renormalization of the gauge coupling.
This term alone is invariant under the ``$U(1)_R$''
transformation. This corresponds to the fact that the
renormalizations of the gauge couplings of ${\cal N}=2$ SUSY gauge
theories are 1-loop exact ({\it i.e.} the corrections from the higher
order loops are absent). 
The second and higher terms in the square bracket 
$\left[ \right.$  $ \left. \right]$ represent non-perturbative corrections. 
Each terms 
are also invariant under the ``$U(1)_R$'' symmetry.
The deformed prepotential (\ref{eq:prepote-deform}) leads to the
K\"{a}hler potential 
\begin{eqnarray}
 K(\Phi,\Phi^\dagger,M_*)=  \Phi^\dagger \Phi &&
 \left[            \left( 
     \frac{1}{g^2(M_*)} 
   - \frac{1}{8\pi^2} \ln \left( \frac{ e \Phi^\dagger \Phi }{M_*^2}\right)
                   \right) 
 \right.          \nonumber \\ 
&& \left.
             +     \left( 
     c_1^{'} e^{-\frac{8\pi^2}{g^2(M_*)} + i \vartheta} 
         \left(\frac{\Phi}{M_*}\right)^2 
   + h.c.          \right) 
            +      \cdots 
 \right],
\label{eq:deform-kahler-1}
\end{eqnarray}
or if canonically normalized,
\begin{equation}
K(\Phi,\Phi^\dagger,M_*)= \Phi^\dagger \Phi \left[ 1 - \frac{g^2(M_*)}{8\pi^2} \ln \left(\frac{e |\Phi|^2}{M_*^2}\right) + \left( c_1' g^2 e^{-\frac{8\pi^2}{g^2(M_*)} + i \vartheta} \left(\frac{\Phi}{M_*}\right)^2 + h.c.\right) + \cdots \right].
\label{eq:deform-kahler-2}
\end{equation}
We can see that the logarithmic correction in the K\"ahler potential is
exactly the 1-loop wave-function renormalization of the $\Phi$ field 
coming from the Yukawa interaction with the Yukawa coupling $\sqrt{2}g$.
This 1-loop renormalization effect for the wave function is already considered in
\cite{Linde-Riotto}. Now, in this ${\cal N}=2$ SUSY model, higher order
(perturbative) wave-function renormalization is absent from the reason
stated above. Therefore, the perturbative radiative corrections are 
already exactly taken into account. 
The non-perturbative correction terms have exponentially suppressed
coefficients $\exp ( - 8\pi^2/g^2 ) \lsim 10^{-6900}$ for $\sqrt{2}g \lsim
10^{-1}$ suggested from the successful inflation as shown later. 
Thus, we may safely neglect the non-perturbative terms. 


We are now at the point to 
discuss radiative corrections from the supergravity sector, since it 
breaks explicitly the ${\cal N}=2$ SUSY. Here, we see that the small 
cut-off scale $M_*$ is a crucial ingredient to suppress the quantum 
effects. 

The total 1-loop corrections to the inflaton potential\footnote{
The total 1-loop corrections to the scalar potential cannot be described 
only by the renormalization of the K\"{a}hler potential. So we discuss the
corrections to the whole scalar potential not to the K\"{a}hler potential.} 
are given by\footnote{This expression includes the 1-loop correction
discussed in the above analysis.}\cite{Gaillard}
\begin{equation}
 V_{\rm 1-loop} = \frac{M_*^2}{32\pi^2} {\rm Str}\left(m^2(\phi)\right)
                       - \frac{\ln (M_*/\mu)}{32\pi^2} {\rm Str}\left(m^4(\phi)\right), 
\end{equation}
where
\begin{eqnarray}
 {\rm Str}(m^2(\phi)) &=& 2(N-5) \hat{V}(\phi) + 2(N-1) (m_{3/2}(\phi))^2 
                         -2 R^i_j e^K W_{;i} W^{\dagger ;j}+ \cdots  
\label{eq:str-2}\\
 {\rm Str}(m^4(\phi)) &=& 2(N+21) \hat{V}^2 + 4(N+5) \hat{V} m_{3/2}^2 +2(N+17)m_{3/2}^4  \nonumber \\
                      & & + 2 e^K \left( W_{;i ;j}W^{\dagger ;i;j} (2\hat{V} + 3 m_{3/2}^2) - 2 R^m_n W_{;m}W^{\dagger ;n}(\hat{V} + m_{3/2}^2) \right) \nonumber \\
                      & & + 2 e^{2K}\left( 2 W^{\dagger ;i} W_{;i;j}
                                             W^{\dagger;j;k} W_{;k}
                                          + 2 W_{;i;j}W^{\dagger;j;k} R^{m \;\; i}_{\; \;n \;\; k} W_{;m}W^{\dagger;n}  + \cdots \right) \nonumber \\
                      & & + \cdots .
\label{eq:str-4}
\end{eqnarray}
Here we take the $M_{\rm Planck}$ to be unity. The $\hat{V}$ denotes the
tree level scalar potential (\ref{eq:tree-sugra}), ``$_{;i}$'' the
covariant derivative, $R_{i\bar{j}k\bar{l}}$ the Riemann curvature
determined from the K\"{a}hler metric, $N$ the number of the
chiral superfields and $ m_{3/2} = e^{\frac{K}{2}} W$.
For details, see \cite{Gaillard}.

Let us show an estimation for the supertraces starting from their
definitions at first.
Masses of order of the Hubble parameter ($H$) are given to 
all scalar fields($\tilde{\chi}$) in addition to the supersymmetric masses $M_\chi $ 
originally exist.
Mass matrices of scalar fields ($\tilde{\chi}$) are schematically\footnote{
The real calculation would not be that simple.
There are also $|M_\chi|^2 |\phi/M_{\rm Planck}|^2$ contributions in the
diagonal elements of this mass matrix. However, these contributions also
exist in fermion mass matrices and both contributions cancel out each
other after we take the supertraces. } 
described as 
\begin{equation}
 (\tilde{\chi}^*,\tilde{\chi})\left( \begin{array}{cc}
	       |M_\chi|^2 + H^2 +\left|H\frac{\phi}{M_{\rm Planck}}\right|^2 &
  	       M_\chi^* H \frac{\phi}{M_{\rm Planck}}  \\
               M_\chi H \frac{\phi^*}{M_{\rm Planck}} & 
	       |M_\chi|^2 + H^2 + \left|H\frac{\phi}{M_{\rm Planck}}\right|^2
		     \end{array}\right)
      \left(\begin{array}{c}
       \tilde{\chi} \\
       \tilde{\chi}^\dagger
	    \end{array}\right).
\end{equation}
Then the supertraces are roughly given by
\begin{eqnarray}
  {\rm Str}(m^2(\phi))  &\sim&   N H^2
               + N H^2 \left| \frac{\phi}{M_{\rm Planck}}\right|^2 + \cdots, 
\label{eq:rough-str-2}\\ 
  {\rm Str}(m^4(\phi))  &\sim& ( N' |M|^2 H^2 + N'' H^4) 
               + ( N' |M|^2 H^2 + N'' H^4  )
                    \left| \frac{\phi}{M_{\rm Planck}}\right|^2  + \cdots ,
\label{eq:rough-str-4}
\end{eqnarray}
where $N'$ is the number of heavy particles, $N''$ that of the 
light particles ($N=N' + N''$) and $M$ denotes masses of heavy particles. 

We find through an explicit calculation with use of the
eqs.(\ref{eq:str-2},\ref{eq:str-4}) that the rough estimation given in
eqs.(\ref{eq:rough-str-2},\ref{eq:rough-str-4}) are indeed correct.
We show here the explicit evaluation of the second term in 
eq.(\ref{eq:str-2}) as an example:
\begin{eqnarray}
 2(N-1) m_{3/2}^2 & = & 2(N-1) e^{\frac{|\phi|^2}{M^2_{\rm Planck}}+\cdots }
  \left| \frac{\sqrt{2}g \mu^2 \phi + \cdots}{M^2_{\rm Planck}} \right|^2  
        \nonumber \\
  & = & 2(N-1) \frac{|\sqrt{2}g \mu^2|^2}{M^2_{\rm Planck}} 
                   \left|\frac{\phi}{M_{\rm Planck}}\right|^2  \nonumber \\
  & & +  2(N-1) \vev{m_{3/2}}^2 \left(
                1 + \left|\frac{\phi}{M_{\rm Planck}}\right|^2  + \cdots 
                               \right) + \cdots.
\end{eqnarray}
The first term gives the $|\phi|^2$ term in eq.(\ref{eq:rough-str-2}):
note that $\sqrt{2}g\mu^2 /M_{\rm Planck} \sim H$ and that the $|\phi|^2$
term from the second term is negligible because $H \gg \vev{m_{3/2}}$
during the inflation. 

The total 1-loop corrections to the inflaton potential is now given by
\begin{eqnarray}
V_{\rm 1-loop} \sim \frac{1}{32\pi^2} N M_*^2 H^2 
& + & \frac{1}{32\pi^2}N\left(\frac{M_*}{M_{\rm Planck}}\right)^2 H^2 |\phi|^2
 \nonumber \\ 
& - & \frac{\ln (M_*/\mu)}{32\pi^2} N' 
\left(\frac{M}{M_{\rm Planck}}\right)^2 H^2 |\phi|^2 + \cdots .
\end{eqnarray}
$M$ denotes heavy particle masses which must be lower than the cut-off
scale $M_*$. Since we know that the number of all particles is large ($N 
\gsim 100$), the 1-loop suppression factor $1/(32\pi^2)$ is compensated
by this large $N$. Then the 1-loop corrections would give a mass term of 
order of the Hubble parameter to the inflaton potential and would
violate the slow roll conditions if the cut-off scale were set to be 
the Planck scale $M_{\rm Planck}$. Furthermore, the perturbation theory 
would be no longer valid and the tree level potential would loose its
meaning.  

In order to suppress the loop corrections sufficiently, we impose
\begin{equation}
 N \left(\frac{M_*}{M_{\rm Planck}}\right)^2 \lsim 1 .
\end{equation}
This means that the cut-off scale $M_*$ should be lower than the
Planck scale as 
\begin{equation}
 M_* \lsim \sqrt{\frac{1}{N}} M_{\rm Planck} \simeq 10^{-1} M_{\rm Planck}. 
\end{equation}

The above hybrid inflation model 
with the 1-loop corrections leads to the same inflaton
potential as that discussed in \cite{Lepto}. 
The Yukawa coupling $\lambda$ in \cite{Lepto} corresponds to the 
$\sqrt{2}g$ in this paper, the coefficient of the mass term  
$3 H^2 |\phi|^2 $ is $k$ in \cite{Lepto} while it is $\sim (1/32\pi^2) 
N (M_*/M_{\rm Planck})^2 $ here. The $\mu^2$ there corresponds to the 
$\sqrt{2}g \mu^2 $ here.
The only difference is the assumption on the reheating processes, but this 
does not lead to a major difference in the allowed parameter region for
the desired inflation. The detailed analysis in \cite{Lepto} shows that
the desired hybrid inflations occur for a wide parameter region of
$\lambda \lsim 1.6 \times 10^{-1}$, $k \lsim 3 \times 10^{-2}$ and $
\mu \sim 10^{13-15}\GEV$, which suggests $ g \lsim 1.1 \times 10^{-1}$, $\mu \simeq
(1-20) \times 10^{15}\GEV$ and $ M_* \lsim 0.2 \times M_{\rm Planck}$ in 
this model.  
\footnote{
One might impose that the inflaton-field value during the inflation is smaller
than the cut-off scale $M_*$.
Then, one finds that the parameter space with $3 \times 10^{-2} \lsim
\sqrt{2}g $ and $ M_* \lsim \vev{\psi}|_{\rm  vac} \sim (0.1 - 2) \times
10^{16} \GEV$ are excluded.
One will see that the Fayet-Iliopoulos parameter $\mu^2$ is always
smaller than the cut-off $M_*^2$ in the remaining allowed region. 
}

The inflaton sector in this scenario has no interaction term with other
sectors (including the standard-model sector) not only in the
superpotential but also in the K\"ahler potential. Thus, the inflaton sector
fields decay only through the supergravity interactions.
Operators relevant to decays of the scalar fields $\psi, \bar{\psi}$
and $\phi$ are
%
\begin{eqnarray}
 V &=& W_\psi \frac{K_{\psi^\dagger}}{M^2_{\rm Planck}} W^\dagger + h.c.+\cdots
      =\sqrt{2}g \phi \vev{\bar{\psi}}\frac{\vev{\psi}}{M^2_{\rm Planck}}
             (y_t h \tilde{q}\tilde{u})^\dagger + h.c. + \cdots , 
                                                     \label{eq:dec-op-phi}\\
 V &=& \exp \left( \frac{K(\psi,\bar{\psi})}{M^2_{\rm Planck}} \right) 
                          |W_{h}|^2 + \cdots     
     = \left(\frac{ 
     2{\rm Re} \left( \vev{K_\psi}\psi + \vev{K_{\bar{\psi}}}\bar{\psi} \right)
                  }
                  {M^2_{\rm Planck}}
       \right)    |y_t \tilde{q}\tilde{u}|^2 + \cdots  ,
                                                    \label{eq:dec-op-R-psi} \\
{\cal L}/ {\rm det}e_\mu^a &=& K_{i\bar{j}} \bar{\chi}^{\bar{j}} 
        i \bar{\sigma}^\mu 
 \left(\partial_\mu \delta^i_k + \Gamma^i_{kl}\partial_\mu \tilde{\chi}^l 
       -\frac{i}{2}\delta^i_k \frac{1}{M^2_{\rm Planck}}
                              {\rm Im}(K_l \partial_\mu \tilde{\chi}^l)
       +\cdots 
 \right)                      \chi^k   \nonumber \\
    &=& -\frac{i}{2}\frac{1}{M^2_{\rm Planck}}{\rm Im}
         \left(\vev{K_\psi} \partial_\mu \psi 
             + \vev{K_{\bar{\psi}}} \partial_\mu \bar{\psi}
         \right)  \left(  \bar{n}i\bar{\sigma}^\mu n + \cdots \right), 
\label{eq:dec-op-I-psi}
\end{eqnarray}
where $h, \tilde{q},\tilde{u}$ denote higgs and squarks, 
$y_t$ the top-quark Yukawa coupling constant, $\tilde{\chi}$ and $\chi$
general scalars and fermions, $n$ a right handed neutrino,
and $\Gamma^i_{kl}$ the Christoffel symbol determined from the K\"ahler 
metric. 
The coherent oscillation of the $\phi$ field decays through
eq.(\ref{eq:dec-op-phi}). We can see that the coherent oscillation 
in the $\psi,\bar{\psi}$ field space is almost parallel to the ${\rm Re} 
(\vev{K_\psi}\psi + \vev{K_{\bar{\psi}}}\bar{\psi})$ direction, which 
decays into radiation of standard-model sector fields through
eq.(\ref{eq:dec-op-R-psi}). 
The decay rates are
\begin{eqnarray}
 \Gamma_\phi & \sim & \frac{1}{8\pi}\left(\sqrt{2}g y_t \left(\frac{\mu}{M_{\rm Planck}}\right)^2\right)^2 m_\phi , \label{eq:decay-phi}\\
 \Gamma_{{\rm R}(\psi)} & \sim & \frac{1}{8\pi}\left(y_t^2 \frac{\mu}{M^2_{\rm Planck}}\right)^2 m_\psi^3 , \label{eq:decay-R-psi} 
\end{eqnarray}
where $m_\phi = m_\psi = 2 g \mu$. Notice that all states in the
inflaton sector form a massive vector multiplet of 
${\cal N}=2$ SUSY, and hence all have the same mass $m = 2 g \mu$.
These two decay rates are almost the same as the decay
rate eq.(34) in \cite{Lepto}, and hence the reheating temperature $T_R$
is given by the Fig.8 in \cite{Lepto}. We see the reheating
temperatures $T_R=10^{4-9}\GEV$ in most of the parameter region, which 
are low enough to avoid the overproduction of gravitinos of $m_{3/2}
\sim 1 \TEV$\cite{gravitino}.
We may invoke the Affleck-Dine type baryogenesis\cite{Aff-Dine} for such 
low reheating temperatures\footnote{
We assume that the K\"ahler potential for quark and lepton
multiplets,$q,l$ is complicated enough to have a minimum at the cut-off
scale, $ q,l \sim M_* $, during the hybrid inflation. }. 
The leptogenesis scenario\cite{Fuku-Yana} 
via thermal production of the right handed neutrinos is marginally
possible\cite{Buch-Plum}.

However, 
the energy density of the coherent oscillation of the inflaton fields 
may be converted into other massive inflaton-sector particles through
parametric resonance effects\cite{KLS} well before the decay to light
particles. 
Not a small fraction of the energy density may now be carried by higgsed
vector bosons, massive Dirac fermions, and scalar field ${\rm
Re}(\vev{\psi}\psi^\dagger-\vev{\bar{\psi}}\bar{\psi^\dagger})$, which
all do not have decay operators. This leads to a horrible matter
dominated universe after the reheating.
If it is the case, we have to introduce small explicit breaking
operators of the 
global ${\cal N}=2$ SUSY in order for those particles to decay. 
We introduce two small breaking interactions: one is in the superpotential,
\begin{equation}
W = \epsilon_1 \Phi H_u H_d,  
\label{eq:breaking1}
\end{equation}
and the other is in the kinetic function of gauge multiplets,
\begin{equation}
 f = \epsilon_2  W^\alpha W_\alpha^{'},
\label{eq:breaking2}
\end{equation} 
 where $W_\alpha$ is the field strength tensor of the $U(1)$ gauge 
multiplet in the inflaton sector and $W_\alpha^{'}$ that of the $U(1)_{B-L}$
 symmetry which is also supposed to be gauged and higgsed(spontaneously 
broken).
Then, the decay rates of the ${\cal N}=1$ massive vector multiplet of the 
 inflaton sector are $ \Gamma \sim (1/8\pi) \epsilon^{2}_2 m $ and 
those of the ${\cal N}=1$ massive chiral multiplets $\Phi$ and 
$(\vev{\Psi}\bar{\Psi}+ \vev{\bar{\Psi}}\Psi)$ 
are not smaller than $\Gamma \sim (1/8\pi) \epsilon_1^2 m$. Here, we have 
assumed that the higgsed $U(1)_{B-L}$ gauge boson mass is lighter than 
the mass $m=m_\psi$ of the inflaton-sector gauge boson.
If we require the reheating temperature after the decays of these
 particles to be larger than the $100\GEV$, we obtain the
breaking parameters $\epsilon_1, \epsilon_2 \gsim
10^{-14}(T_R/100\GEV)$. 
The effect of these small breakings to the inflaton potential
 is negligible. 

One can identify the $U(1)$ gauge symmetry of this hybrid inflation
sector with the $U(1)_{B-L}$
symmetry itself. In this case, one must impose ${\cal N}=2$ SUSY on the whole
standard model sector, as was studied in \cite{Polonski-Su}. Then we do
not have to introduce the above breaking operators
(\ref{eq:breaking1},\ref{eq:breaking2}) to
make reheating process successful.
 
In this paper, we have proposed a use of ${\cal N}=2$ SUSY to control the 
tree-level K\"ahler potential for the inflaton $\varphi$ and
constructed a hybrid inflation model based on the ${\cal N}=2$ SUSY
$U(1)$ gauge theory. We have found that the K\"ahler potential for the
inflaton superfield $\Phi$ is fixed as the minimal form $K=\Phi^\dagger
\Phi$ by the ${\cal N}=2$ SUSY and the $U(1)_R$ symmetry. Thus, the flat 
potential at the tree level is a consequence of the symmetries. In
this model, the radiative corrections from the supergravity interactions 
are sufficiently suppressed by the low cut-off scale $M_* \lsim 10^{-1}
M_{\rm Planck}$, and the model turns out to be almost the same as that
investigated intensively in \cite{Lepto}. The inflation potential 
in our model is determined by 
three basic parameters, namely the $U(1)$ gauge coupling constant $g$,
Fayet-Iliopoulos parameter $\mu^2$ and the cut-off scale $M_*$.


The global ${\cal N}=2$ SUSY in this model forbids all mixed terms in
the K\"ahler and superpotentials between the inflaton sector and 
all other sectors. The reheating takes place only through the 
supergravity interactions with $1/M^2_{\rm Planck}$ suppressed
amplitudes. If the parametric resonance effect transfers some part of 
the coherent oscillation energy to various inflaton-sector particles, 
then we have to introduce small breaking operators of the ${\cal N}=2$ 
SUSY in order to make reheating process sufficiently fast. 

We have also assumed that the K\"ahler potential is determined only by the
cut-off scale $M_*$ and not by $M_{\rm Planck}$. This assumption may be 
based on the picture that the cut-off scale $M_*$ is truly the 
fundamental scale of the theory, which is lower than the $M_{\rm
Planck}$.
Here, $M_{\rm Planck}^2$ may be merely an effective scale appearing in 
the effective 4 dimensional gravity below the fundamental scale.
The physics underlying our assumption is under investigation.

\section*{Acknowledgment}
Authors thank Masahiro Kawasaki for discussion.
This work is partially supported by ``Priority Area: Supersymmetry and
Unified Theory of Elementary Particles(No. 707)''(T.Y.).  T.W. thanks
the Japan Society for the Promotion of Science.

\end{document}